\documentclass{aa}
\usepackage{natbib}
\usepackage[dvips,final]{graphics}
\usepackage{psfrag}

\psfrag{l}{$\lambda$}
\psfrag{W}{$\Omega$}
\psfrag{p}{$p$}
\psfrag{F}{$F$}
\psfrag{H}{$($}
\psfrag{D}{$D$}
\psfrag{\310}{$|$}
\psfrag{L}{$)$}
\psfrag{=}{$=$}
\psfrag{,}{$,$}
\psfrag{0.3}{$0.3$}

\begin{document}

\thesaurus{02(12.07.1; 12.03.4; 12.03.3)}

\title{
Gravitational lensing statistics with extragalactic surveys
}

\subtitle{
II. Analysis of the Jodrell Bank-VLA Astrometric Survey
}

\author{
Phillip Helbig\inst{1}
\and
Daniel Marlow\inst{1}\thanks{\emph{Present Address:}
University of Pennsylvania, 
Dept.~of Physics and Astronomy, 
209 S.~$33^\mathrm{rd}$ Street, 
Philadelphia, PA 19104-6396,
U.S.A.
}
\and
Ralf Quast\inst{2} 
\and 
Peter N.~Wilkinson\inst{1}
\and
Ian W. A. Browne\inst{1}
\and
L.~V.~E.~Koopmans\inst{3}
} 

\institute{
University of Manchester, 
Nuffield Radio Astronomy Laboratories, 
Jodrell Bank, 
Macclesfield,
Cheshire SK11 9DL, 
UK
\and 
Universit{\"a}t Hamburg, 
Hamburger Sternwarte, 
Gojenbergsweg 112, 
D-21029 Hamburg, 
Germany
\and 
University of Groningen,
Kapteyn Astronomical Institute, 
Postbus 800, 
NL-9700 AV Groningen,
The Netherlands
}

\date{Received December 2, 1998; accepted January 12, 1999}

\offprints{P.~Helbig}
\mail{p.helbig@jb.man.ac.uk}

\authorrunning{P.~Helbig et al.}
\titlerunning{
Gravitational lensing statistics with extragalactic surveys. II
}

\maketitle

\begin{abstract}
We present constraints on the cosmological constant $\lambda_{0}$ from
gravitational lensing statistics of the Jodrell Bank-VLA Astrometric
Survey (JVAS).  Although this is the largest gravitational lens survey
which has been analysed, cosmological constraints are only comparable to
those from optical surveys.  This is due to the fact that the median
source redshifts of JVAS are lower, which leads to both relatively fewer
lenses in the survey and a weaker dependence on the cosmological
parameters.  Although more approximations have to be made than is the
case for optical surveys, the consistency of the results with those from
optical gravitational lens surveys and other cosmological tests indicate
that this is not a major source of uncertainty in the results.  However,
joint constraints from a combination of radio and optical data are much
tighter.  Thus, a similar analysis of the much larger Cosmic Lens
All-Sky Survey should provide even tighter constraints on the
cosmological constant, especially when combined with data from optical
lens surveys. 

At 95\% confidence, our lower and upper limits on
$\lambda_{0}-\Omega_{0}$, using the JVAS lensing statistics information
alone, are respectively $-2.69$ and $0.68$.  For a flat universe, these
correspond to lower and upper limits on $\lambda_{0}$ of respectively
$-0.85$ and $0.84$.  Using the combination of JVAS lensing statistics and
lensing statistics from the literature as discussed in
\citet{RQuastPHelbig99a} the corresponding $\lambda_{0}-\Omega_{0}$
values are $-1.78$ and $0.27$.  For a flat universe, these correspond to
lower and upper limits on $\lambda_{0}$ of respectively $-0.39$ and
$0.64$. 

\end{abstract}

\section{Introduction}

The use of gravitational lensing statistics as a cosmological tool was
first considered in detail by \citet{ETurnerOG84a}; the influence of the
cosmological constant was investigated thoroughly by
\citet{MFukugitaFKT92a}, building on the work of \citet{ETurner90a} and
\citet{MFukugitaFK90a}.  \citet[and references therein]{CKochanek96a}
and, more recently, \citet[hereafter FKM]{EFalcoKM98a} have laid the
groundwork for using gravitational lensing statistics for the detailed
analysis of extragalactic surveys.  \citet[hereafter
Paper~I]{RQuastPHelbig99a} reanalysed optical surveys from the
literature, for the first time exploring a range of the
$\lambda_{0}$-$\Omega_{0}$ parameter space large enough to enable a
comparison with other cosmological tests.  Here, we use the formalism
outlined in Paper~I to analyse the Jodrell Bank-VLA Astrometric Survey
(JVAS), the largest completed gravitational lens survey to date. 

Radio surveys offer several advantages over optical surveys (see, e.g.,
FKM): one doesn't have to worry about systematic errors due to
extinction or a lens galaxy of apparent brightness comparable to that of
the lensed images of the source, the resolution (of followup
observations if not of the survey proper) is much smaller than the
typical image separation, parent catalogues in the form of large-area
surveys exist from which unbiased samples can be selected and relatively
easily observed.  Disadvantages in the radio are due to our relatively
poor knowledge of the flux density-dependent redshift distribution or
equivalently the redshift-dependent number-magnitude relation. 

For a description of our method see Paper~I.  The plan of this paper is
as follows.  Sect.~\ref{JVAS} describes the JVAS gravitational lens
survey.  In Sect.~\ref{calculations} we describe the calculations we
have done based on the JVAS data.  Sect.~\ref{results} presents our
results, using both the JVAS data alone and in combination with the
results from the optical surveys analysed in Paper~I.  Finally in
Sect.~\ref{conclusions} we compare our results to those of Paper~I and
present our conclusions and our prognosis for the analysis of future
large surveys such as CLASS.

\section{The JVAS Gravitational Lens Survey}
\label{JVAS}

\subsection{The sample}

The Jodrell Bank-VLA Astrometric Survey (JVAS) is a survey for
flat-spectrum radio sources with a flux density greater than 200\,mJy at
5\,GHz.  Flat-spectrum radio sources are likely to be compact, thus
making it easy to recognise the lensing morphology.  In addition, they
are likely to be variable, making it possible to determine $H_{0}$ by
measuring the time delay between the lensed images.  (See
\citet{ABiggsBHKWP99a} for the description of a time delay measurement
in a JVAS gravitational lens system.)  JVAS is also a survey for MERLIN
phase-reference sources and as such is described in
\citet{APatnaikBWW92a}, \citet{IBRownePWW98a} and
\citet{PWilkinsonBPWS98a}.  JVAS as a gravitational lens survey, the lens
candidate selection, followup process, confirmation criteria and a
discussion of the JVAS gravitational lenses is described in detail in
\citet{LKingBMPW99a} \citep[see also][]{LKingIBrowne96a}. 

In order to have a parent sample which is as large as possible and as
cleanly defined as practical, our ``JVAS gravitational lens survey
sample'' is slightly different than the ``JVAS phase-reference calibrator
sample''.  For the former, the source must be a point source and must
have a good starting position (so that the observation was correctly
pointed) while its precise spectral index is not important.  For the
latter, only the spectral index is important, as the source can be
slightly resolved or the observation can be less than perfectly pointed.
Thus, the JVAS astrometric sample
\citep{APatnaikBWW92a,IBRownePWW98a,PWilkinsonBPWS98a} contains 2144
sources.  To these must be added 103 sources which were too resolved to
be used as phase calibrators and 61 sources which had bad starting
positions (thus the observations were too badly pointed to be useful for
the astrometric sample), bringing the total to 2308.  This formed our
gravitational lens sample, since these additional sources were also
searched for gravitational lenses \citep{LKingBMPW99a} (none were found
meeting the JVAS selection criteria).

\subsection{The lenses}

We have used the gravitational lens systems in Table~\ref{ta:lenses} in 
this analysis.  
\begin{table*}
\caption[]{JVAS lenses used in this analysis.  Of the information in the 
table, for this analysis we use only the source redshift $z_{\mathrm{s}}$
and the image separation $\Delta\theta$}
\label{ta:lenses}
\begin{tabular*}{\textwidth}{@{\extracolsep{\fill}}llllll}
\hline
\hline
Name & 
\# images & 
$\Delta\theta ['']$ & 
$z_{\mathrm{l}}$ & 
$z_{\mathrm{s}}$ & 
lens galaxy \\
\hline
\object{B0218$+$357}   &                
2 + ring               &                
0.334                  &                
0.6847                 &                
0.96                   &                
spiral                 \\               
\object{MG0414$+$054}  &                
4                      &                
2.09                   &                
0.9584                 &                
2.639                  &                
elliptical             \\               
\object{B1030$+$074}   &                
2                      &                
1.56                   &                
0.599                  &                
1.535                  &                
spiral                 \\               
\object{B1422$+$231}   &                
4                      &                
1.28                   &                
0.337                  &                
3.62                   &                
\textbf{?}             \\               
\hline
\end{tabular*}
\end{table*}
The JVAS lens \object{B1938$+$666} \citep{LKingJBBBdBFKMNW98a} was not
included because it is not formally a part of the sample, having a too
steep spectral index and having been recognised on the basis of a lensed
extended source as opposed to lensed compact components.  Also, the JVAS
lens \object{B2114$+$022} \citep{PAugustoBWJFM99a} was not included
because it is not a single-galaxy lens system.

\section{Calculations}
\label{calculations}

A major difference between the analysis of an optical survey (see
Paper~I and references therein) and a radio survey is that in the latter
one does not know the redshifts of all the unlensed sources.  One can
still use the formalism of Paper I, however, substituting for the
non-lensed objects in the sample a subsample with known redshifts,
multiplying the logarithm of this contribution from the non-lenses to
the likelihood by the ratio of the size of the parent sample to that of
the subsample.  Alternatively, one can take the redshifts from a sample
selected according to the same criteria, assigning these randomly to
objects in (a subsample of) the parent sample for a similar flux density
range.  Similarly, one does not know the number-magnitude relation for
the sample and for its extension to fainter flux densities (needed to
allow for the lens amplification).  Again, this can be estimated from
either a subsample (through extrapolation) or from another sample
selected according to the same criteria (either through extrapolation or
by having a fainter flux density limit in this other sample; in the
latter case obviously the selection criteria should be identical to that
of the original sample except for the lower flux density limit). 

For this analysis, due to the paucity of the observational data, we have
made rather stark assumptions: the redshift distribution of the sample
is assumed to be identical to that of the CJF sample
\citep{GTaylorVRPHW96a}, independent of flux density, and the
number-magnitude relation is assumed to be identical to that of CLASS
\citep[Cosmic Lens All-Sky Survey,][]{SMyersetal99a}, independent of
redshift. 

Otherwise, we have followed the procedure outlined in Paper~I,
calculating the a priori likelihood of obtaining the observational data
as a function of $\lambda_{0}$ and $\Omega_{0}$ and the a posteriori
likelihood for the three different choices of prior information used in
Paper I.  We present results both for the JVAS lens survey and for the
combination of the JVAS results with those from the optical surveys
analysed in Paper~I.

\section{Results and discussion}
\label{results}

\begin{figure*}
\resizebox{0.375\textwidth}{!}{\includegraphics{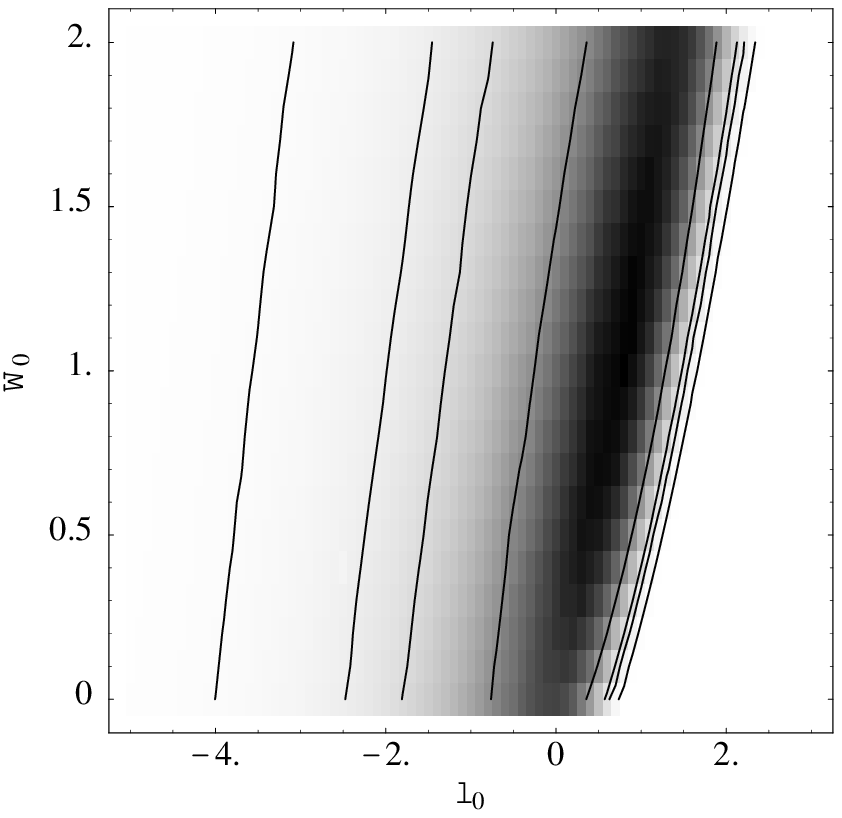}}
\hfill
\resizebox{0.375\textwidth}{!}{\includegraphics{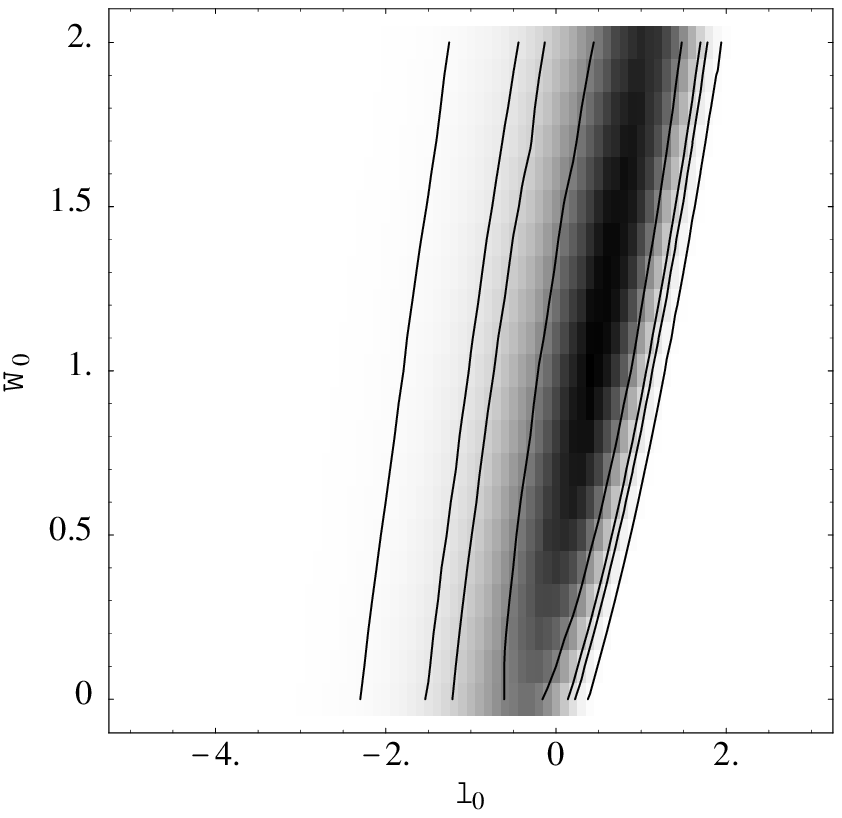}}
\caption[]{\emph{Left panel:} The likelihood function
$p(D|\lambda_0,\Omega_0,\vec{\xi}_0)$ based on the JVAS lens sample. 
All nuisance parameters are assumed to take precisely their mean values.
The pixel grey level is directly proportional to the likelihood ratio,
darker pixels reflect higher ratios.  The pixel size reflects the
resolution of our numerical computations.  The contours mark the
boundaries of the minimum $0.68$, $0.90$, $0.95$ and $0.99$ confidence
regions for the parameters $\lambda_0$ and $\Omega_0$.  \emph{Right
panel:} Exactly the same as the left panel, but the joint likelihood
from the JVAS lens sample and the optical samples from
\citet[Paper~I]{RQuastPHelbig99a}} 
\label{fi:likelihood}
\end{figure*}
\begin{figure*}
\resizebox{0.375\textwidth}{!}{\includegraphics{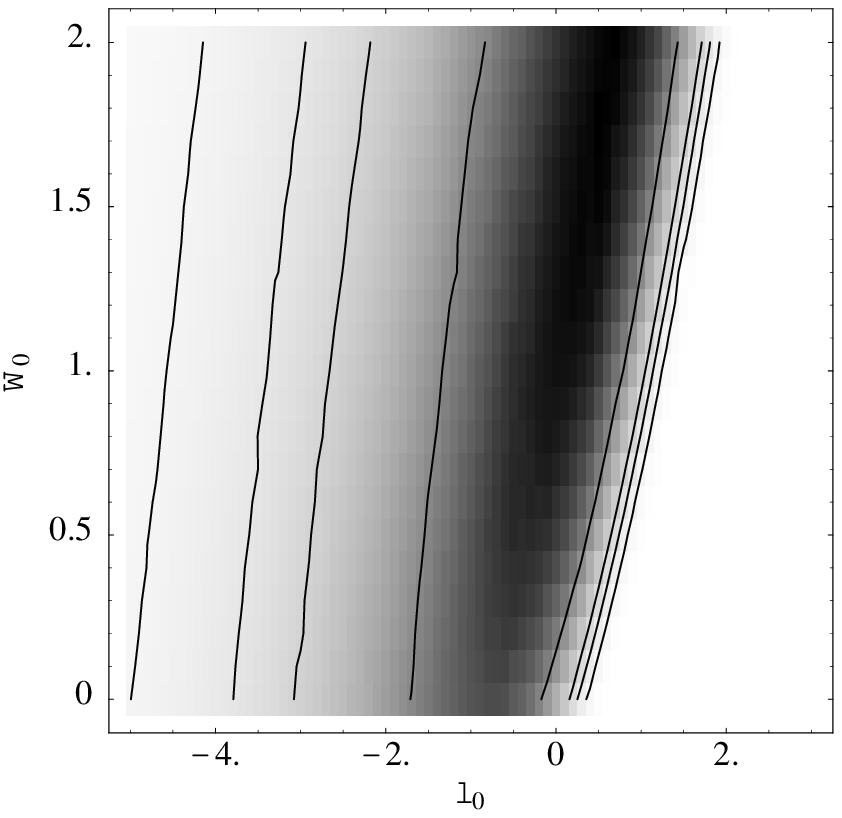}} 
\hfill
\resizebox{0.375\textwidth}{!}{\includegraphics{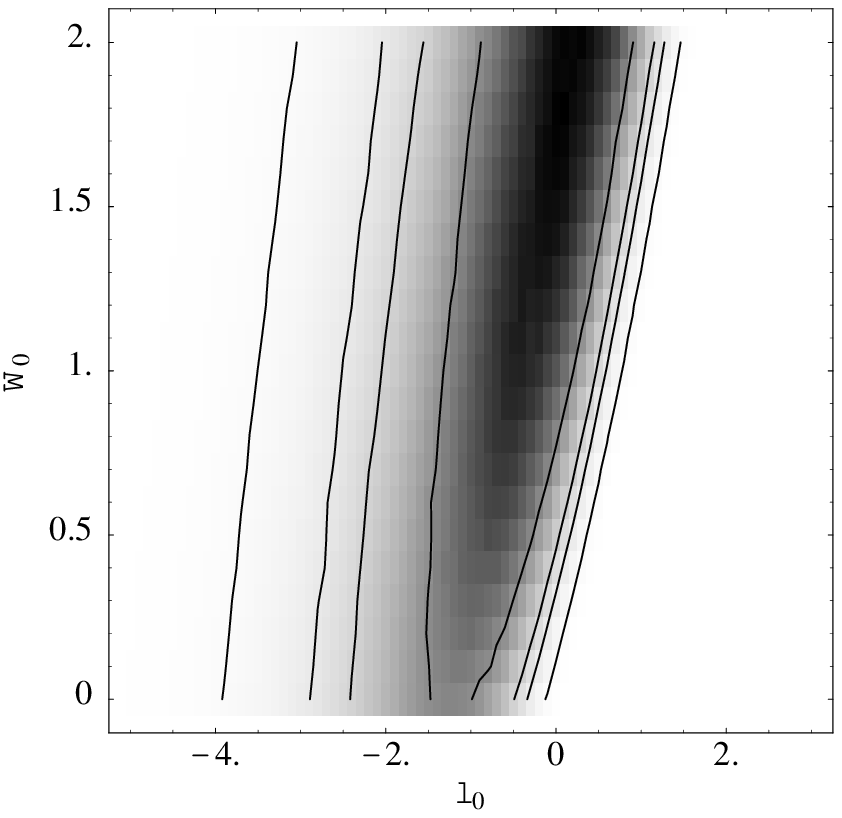}}
\caption[]{Exactly the same as Fig.~\ref{fi:likelihood}, but the
parameter $n_{\mathrm{e}}$ is increased by two standard deviations. 
This parameter, the normalisation of the luminosity function of the lens
galaxies, is one of the more uncertain input parameters, thus one can
get a rough estimate of the overall uncertainty by comparing this figure
and Fig.~\ref{fi:likelihood}.  See the discussion in Paper~I}
\label{fi:xlikelihood}
\end{figure*}

The left panel of Fig.~\ref{fi:likelihood} shows the constraints on the
cosmological parameters $\lambda_{0}$ and $\Omega_{0}$ based only on the
information obtained from the JVAS lens statistics, while the right
panel shows the joint constraints from the JVAS lens sample and the
optical samples from Paper~I.  Fig.~\ref{fi:xlikelihood} is identical
except that one of 
the input parameters, the normalisation of the galaxy luminosity 
function, was increased by two standard deviations.  This gives an idea 
of the magnitude of systematic uncertainties.  (See the discussion in 
Paper~I.) 

\begin{figure*}
\resizebox{0.375\textwidth}{!}{\includegraphics{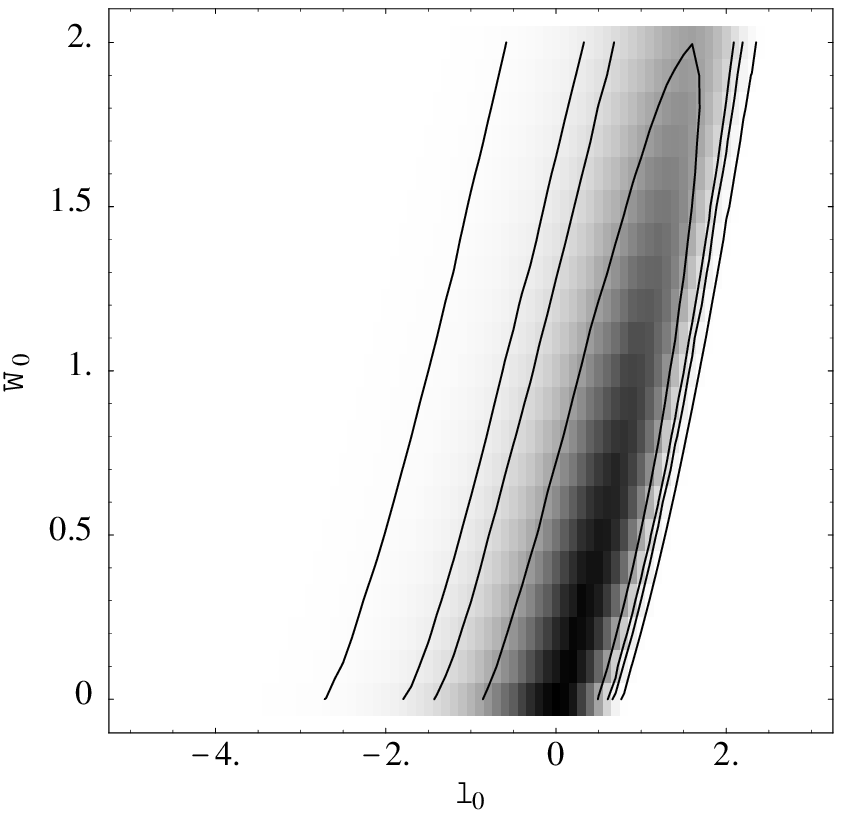}}
\hfill
\resizebox{0.375\textwidth}{!}{\includegraphics{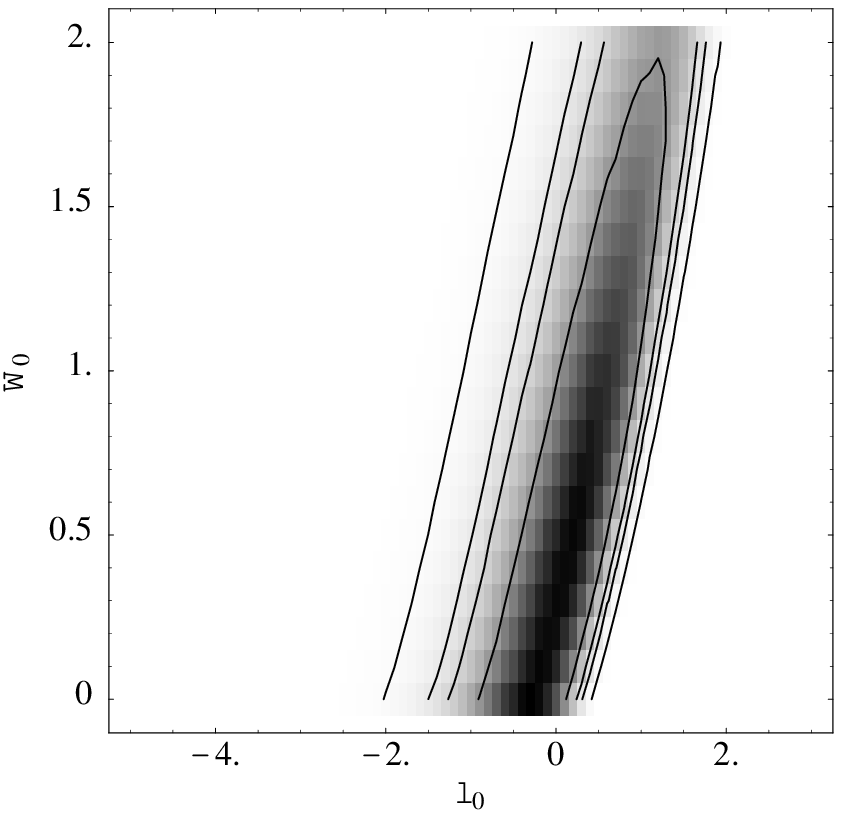}}

\noindent
\resizebox{0.375\textwidth}{!}{\includegraphics{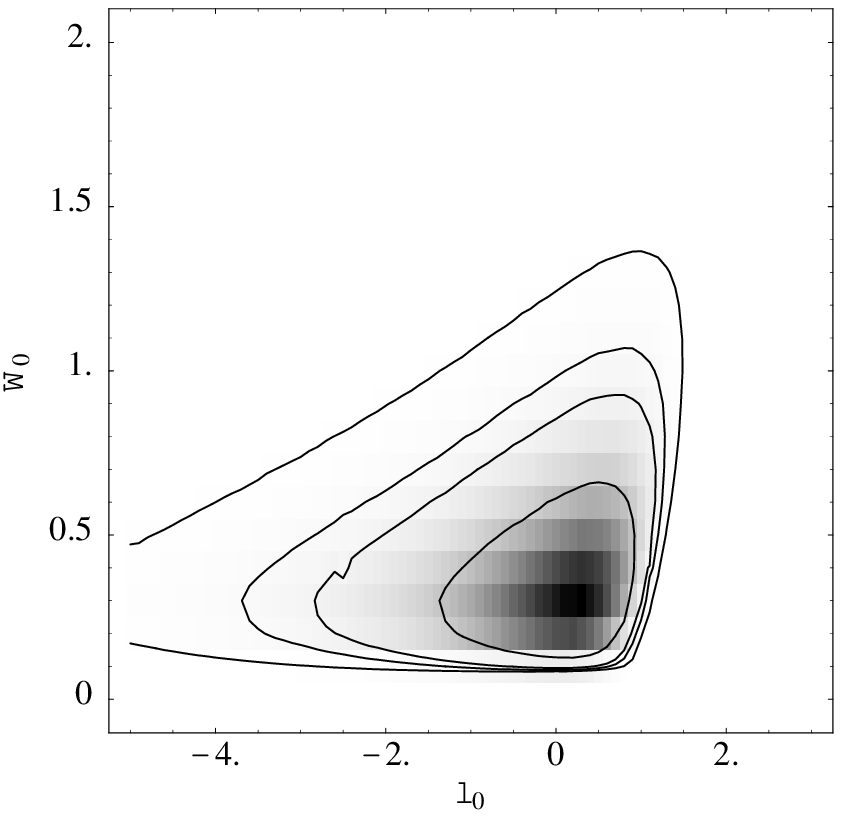}}
\hfill
\resizebox{0.375\textwidth}{!}{\includegraphics{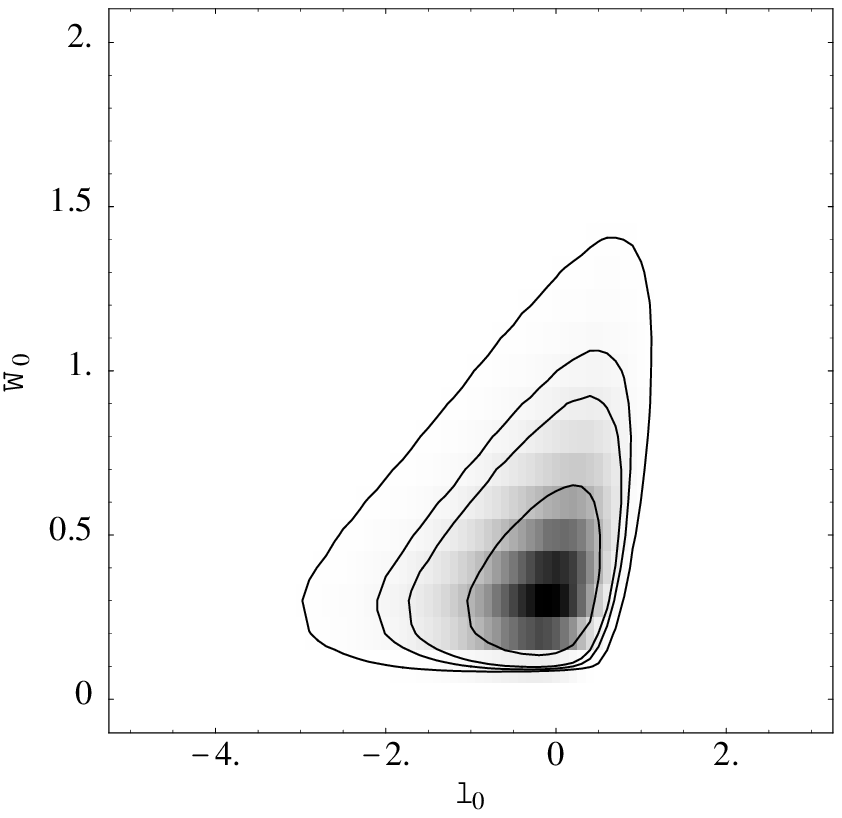}}

\noindent
\resizebox{0.375\textwidth}{!}{\includegraphics{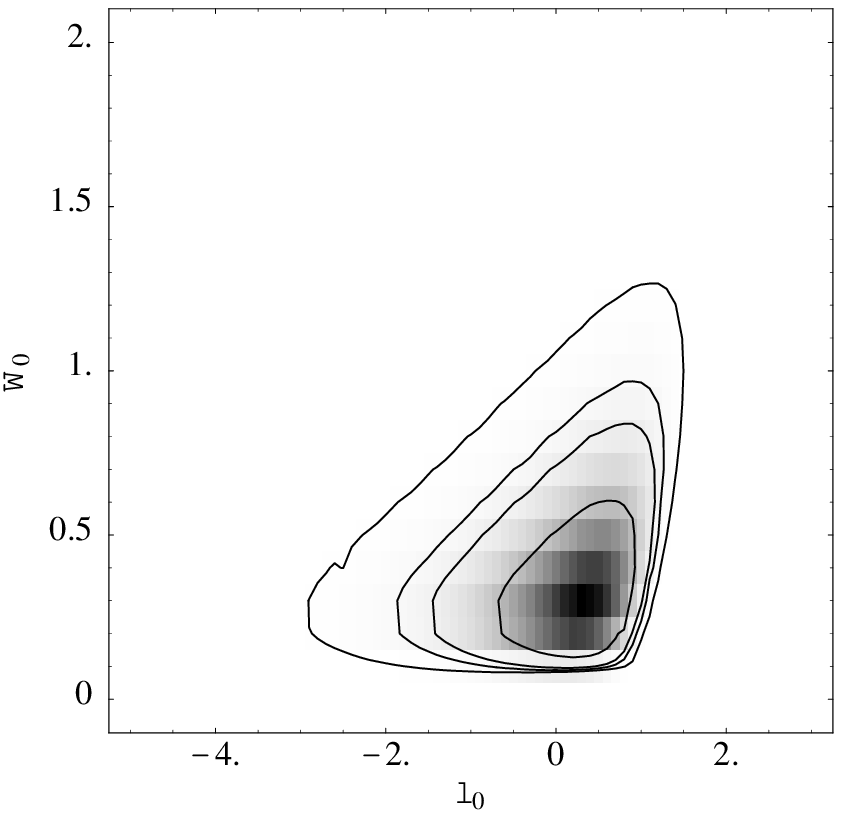}}
\hfill
\resizebox{0.375\textwidth}{!}{\includegraphics{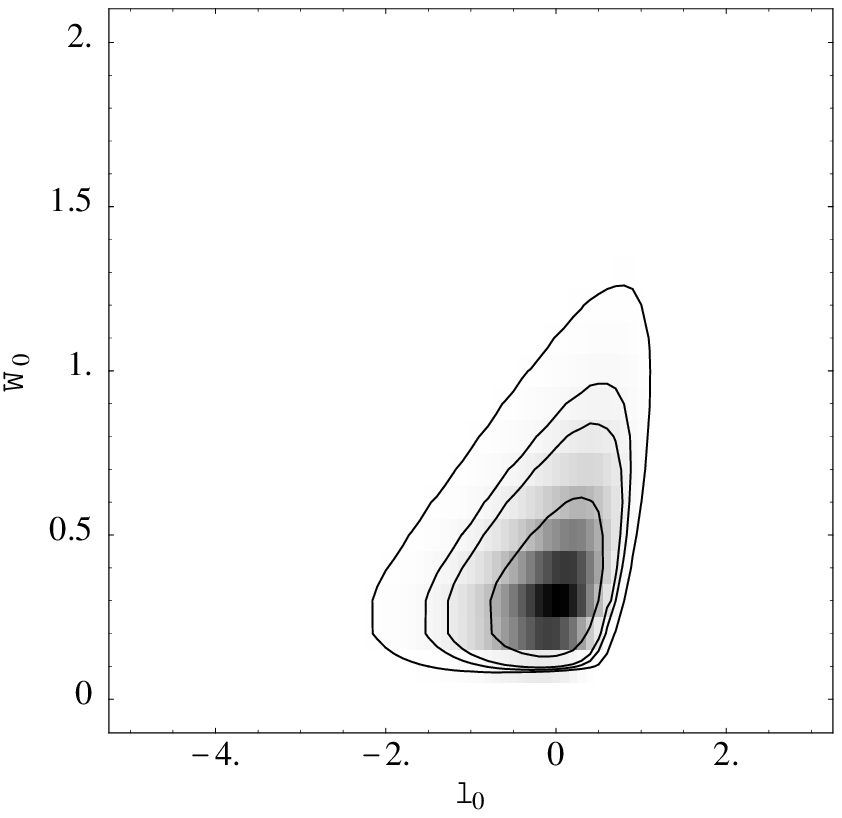}}
\caption[]{\emph{Left column:} The posterior probability density
functions $p_1(\lambda_0,\Omega_0|D)$ (top panel),
$p_2(\lambda_0,\Omega_0|D)$ (middle panel) and
$p_3(\lambda_0,\Omega_0|D)$ (bottom panel).  All nuisance parameters are
assumed to take precisely their mean values.  The pixel grey level is
directly proportional to the likelihood ratio, darker pixels reflect
higher ratios.  The pixel size reflects the resolution of our numerical
computations.  The contours mark the boundaries of the minimum $0.68$,
$0.90$, $0.95$ and $0.99$ confidence regions for the parameters
$\lambda_0$ and $\Omega_0$.  The respective amounts of information
obtained from our sample data are $I_1=1.42$, $I_2=1.32$ and $I_3=1.45$.
\emph{Right column:} Exactly the same as the left panel, but the joint
likelihood from the JVAS lens sample and the optical samples from
\citet{RQuastPHelbig99a}.  The respective amounts of information
obtained from our joint sample data are $1.98$, $1.95$ and $1.96$.  See
Paper~I for definitions} 
\label{fi:posterior}
\end{figure*}
The left plot in the top row of Fig.~\ref{fi:posterior} shows the joint
likelihood of our lensing statistics analysis and that obtained by using
conservative estimates for $H_{0}$ and the age of the universe (see
Paper~I).  Although neither method alone sets useful constraints on
$\Omega_{0}$, their combination does, since the constraint from 
$H_{0}$ and the age of the universe
only allows large values of $\Omega_{0}$ for $\lambda_{0}$ values which
are excluded by lens statistics.  Even though the 68\% confidence
contour still allows almost the entire $\Omega_{0}$ range, it is obvious
from the grey scale that much lower values of $\Omega_{0}$ are favoured
by the joint constraints.  The upper limit on $\lambda_{0}$ changes only
slightly while, as is to be expected, the lower limit becomes tighter.
Right plot: exactly the same, but including optical constraints from
Paper~I.  The upper limits on $\lambda_{0}$ decrease slightly, while the
lower limits improve considerably.  The latter is probably due to the
fact that, in addition to just using more data the JVAS sources are at
significantly different redshifts than those from the optical surveys
analysed in Paper~I (the JVAS sources are generally at lower redshift). 
The former is consistent with the slightly higher optical depth for
radio surveys found by FKM and will be discussed more below. 

The middle row of Fig.~\ref{fi:posterior} shows the effect of including
our prior information on $\Omega_{0}$ (see Paper I).  As is to be
expected, (for both the JVAS and combination data sets) lower values of
$\Omega_{0}$ are favoured.  This has the side effect of weakening our
lower limit on $\lambda_{0}$ (though only slightly affecting the upper
limit).  This should not be regarded as a weakness, however, since
including prior information for $\lambda_{0}$ and $\Omega_{0}$ from the
constraint from $H_{0}$ and the age of the universe as well as for
$\Omega_{0}$ itself, as illustrated in the bottom row of
Fig.~\ref{fi:posterior}, tightens the lower limit again (without
appreciably affecting the upper limit). 

We believe that the right plot of the bottom row of
Fig.~\ref{fi:posterior} represents very robust constraints in the
$\lambda_{0}$-$\Omega_{0}$ plane.  The upper limits on $\lambda_{0}$
come from gravitational lensing statistics, which, due to the extremely
rapid increase in the optical depth for larger values of $\lambda_{0}$,
are quite robust and relatively insensitive to uncertainties in the
input data (cf.~Fig.~\ref{fi:xlikelihood} and the discussion of the
effect of changing the most uncertain input parameter by 2\,$\sigma$ in
Paper~I) as well as to the prior information used (compare the upper,
lower and middle rows of Fig.~\ref{fi:posterior}).  The combination of
data from JVAS and optical surveys leads to much tighter lower limits on
$\lambda_{0}$ than using either alone.  The upper and lower limits on
$\Omega_{0}$ are based on a number of different methods and appear to be
quite robust (see Paper I).  The combination of the relatively secure
knowledge of $H_{0}$ and the age of the universe combine with lens
statistics to produce a good lower limit on $\lambda_{0}$, although this
is to some extent still subject to the caveats mentioned above. 

\begin{figure*}
\resizebox{0.375\textwidth}{!}{\includegraphics{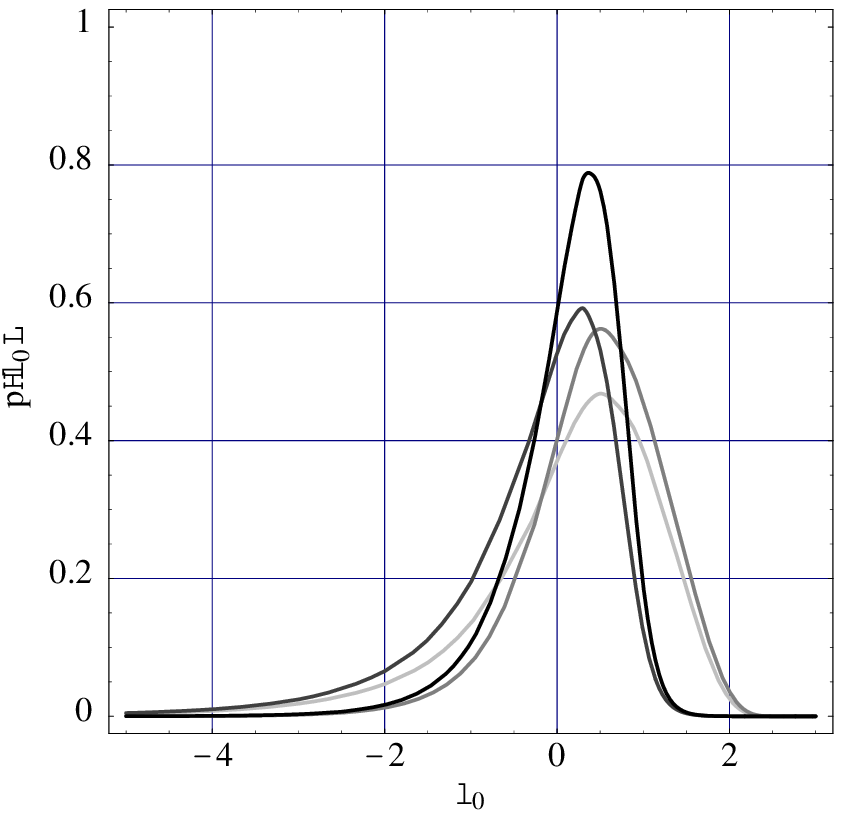}}
\hfill
\resizebox{0.375\textwidth}{!}{\includegraphics{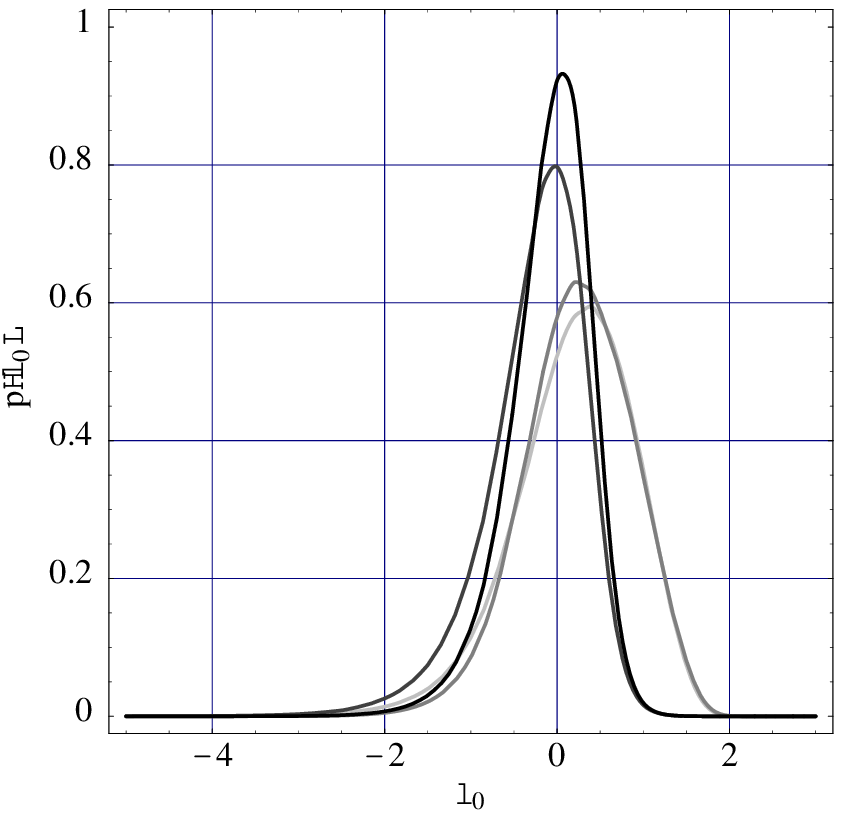}}

\noindent
\resizebox{0.375\textwidth}{!}{\includegraphics{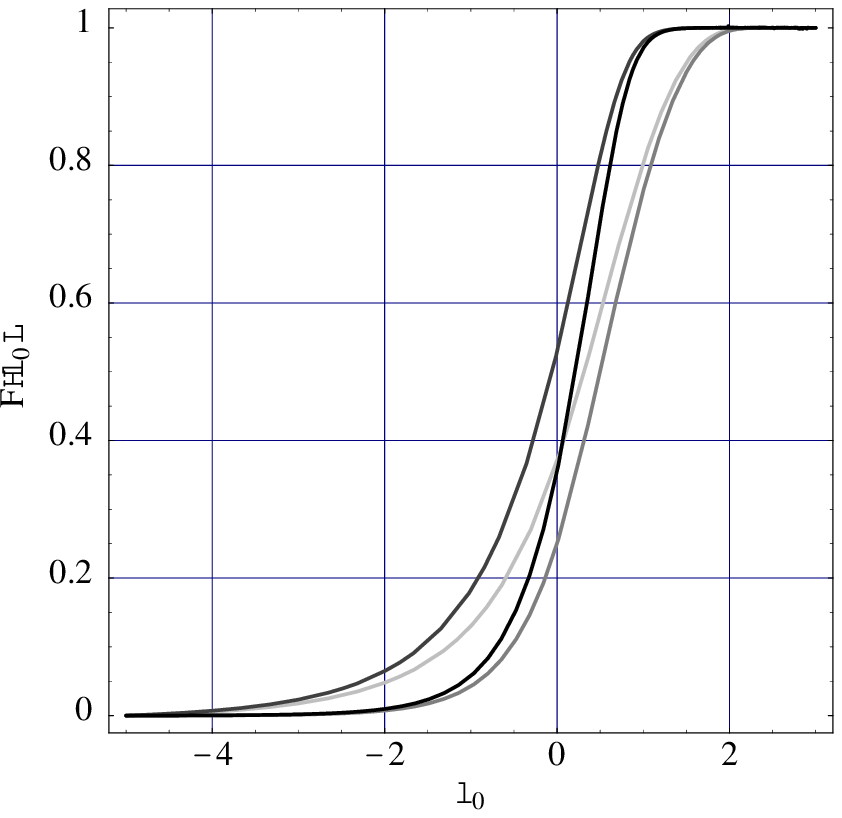}}
\hfill
\resizebox{0.375\textwidth}{!}{\includegraphics{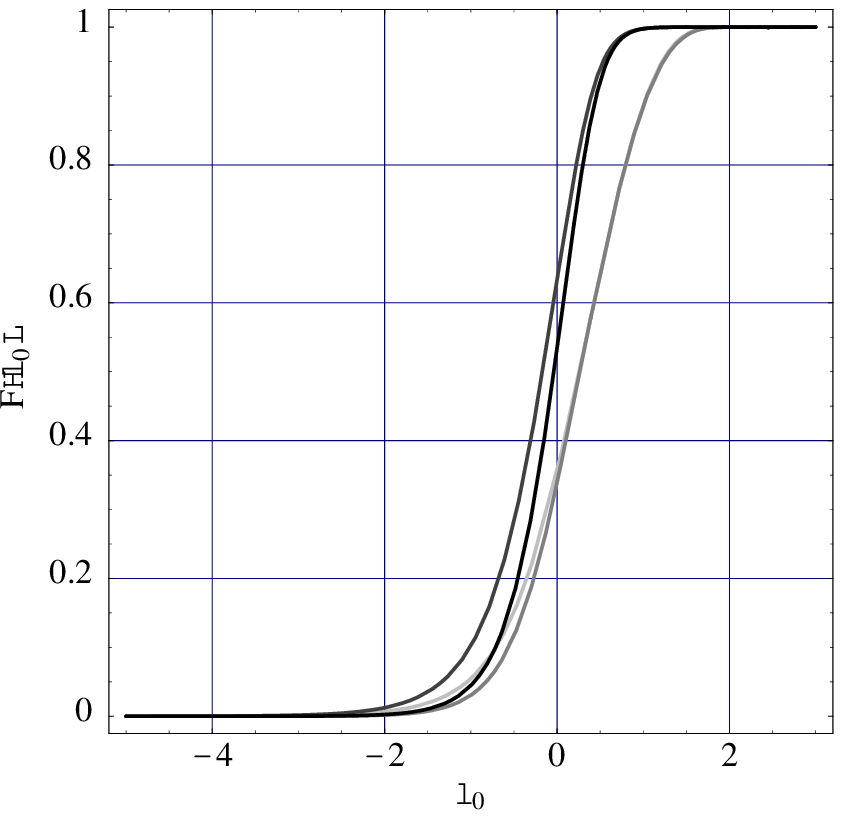}}
\caption[]{\emph{Left column:} The top panel shows the normalised
marginal likelihood function $p(\lambda_0|D)$ (light gray curve) and the
marginal posterior probability density functions $p_1(D|\lambda_0)$
(medium gray curve), $p_2(D|\lambda_0)$ (dark gray curve) and
$p_3(D|\lambda_0)$ (black curve) derived from the JVAS analysis.  All
nuisance parameters are assumed to take precisely their mean values. 
The bottom panel shows the respective cumulative distribution functions.
\emph{Right column:} Exactly the same as the left panel, but the joint
likelihood from the JVAS lens sample and the optical samples from
\citet{RQuastPHelbig99a}} 
\label{fi:marginal}
\end{figure*}

If one is interested in the allowed range of $\lambda_{0}$, one can
marginalise over $\Omega_{0}$ to obtain a probability distribution for
$\lambda_{0}$.  This is illustrated in Fig.~\ref{fi:marginal} 
and Table~\ref{ta:results}.
\begin{table*}
\caption[]{Marginal mean values, standard deviations and $0.95$
confidence intervals for the parameter~$\lambda_0$ on the basis of the
marginal distributions shown in the top row of Fig.~\ref{fi:marginal}} 
\label{ta:results}
\begin{tabular*}{\linewidth}{@{\extracolsep{\fill}}llccccc}
\hline
\hline
Sample & Distribution & Mean & standard deviation  &
\multicolumn{2}{l}{95\% c.l.~range}  & information \\
\hline
JVAS & $p(D|\lambda_0)$   & $0.13$  & $1.08$ & $-2.08$ & $1.91$  & \\
JVAS & $p_1(\lambda_0|D)$ & $0.44$  & $0.77$ & $-1.05$ & $1.87$ & $1.42$  \\
JVAS & $p_2(\lambda_0|D)$ & $-0.29$  & $0.98$ & $-2.38$ & $1.17$ & $1.32$ \\
JVAS & $p_3(\lambda_0|D)$ & $0.11$  & $0.64$ & $-1.20$ & $1.16$ & $1.45$  \\
joint & $p(D|\lambda_0)$   & $0.19$  & $0.70$ & $-1.17$ & $1.48$  & \\
joint & $p_1(\lambda_0|D)$ & $0.24$  & $0.63$ & $-0.96$ & $1.46$ & $1.98$  \\
joint & $p_2(\lambda_0|D)$ & $-0.25$  & $0.59$ & $-1.46$ & $0.77$ & $1.95$ \\
joint & $p_3(\lambda_0|D)$ & $-0.09$  & $0.48$ & $-1.08$ & $0.77$ & $1.96$  \\
\hline
\end{tabular*}
\end{table*}

The comparison values from this work corresponding to those in Tables~3
and 4 of Paper~I are presented in Tables~\ref{ta:specialo} and
\ref{ta:specialk}. 
\begin{table*}
\caption[]{Mean values and ranges for assorted confidence levels for the
parameter $\lambda_{0}$ for our a priori and various a posteriori
likelihoods from this work for $\Omega_{0}=0.3$.  This should be
compared to Table~3 in Paper~I} 
\label{ta:specialo}
\begin{tabular*}{\linewidth}{@{\extracolsep{\fill}}lrrrrrrrr}
\hline
\hline
Cosmological test&
\multicolumn{2}{c}{68\% c.l.~range}  &
\multicolumn{2}{c}{90\% c.l.~range}  &
\multicolumn{2}{c}{95\% c.l.~range}  &
\multicolumn{2}{c}{99\% c.l.~range}  \\
\hline
JVAS, $p(D|\lambda_0)$ &
$-0.66$ & $0.72$ & $-1.68$ & $0.87$ & $-2.36$ & $0.96$ & $-3.91$ & $1.08$ \\
JVAS, $p_1(\lambda_0|D)$ &
$-0.44$ & $0.80$ & $-1.00$ & $0.92$ & $-1.38$ & $1.00$ & $-2.27$ & $1.09$ \\
JVAS, $p_2(\lambda_0|D)$ &
$-1.38$ & $0.86$ & $-2.81$ & $1.00$ & $-3.70$ & $1.06$ & $<-5.00$ & $1.15$ \\
JVAS, $p_3(\lambda_0|D)$ &
$-0.69$ & $0.86$ & $-1.45$ & $0.99$ & $-1.89$ & $1.03$ & $-2.91$ & $1.15$ \\
JVAS \& optical, $p(D|\lambda_0)$ &
$-0.54$ & $0.26$ & $-1.08$ & $0.44$ & $-1.41$ & $0.54$ & $-2.15$ & $0.70$ \\
JVAS \& optical, $p_1(\lambda_0|D)$ &
$-0.63$ & $0.40$ & $-0.95$ & $0.53$ & $-1.18$ & $0.62$ & $-1.72$ & $0.73$ \\
JVAS \& optical, $p_2(\lambda_0|D)$ &
$-1.02$ & $0.44$ & $-1.72$ & $0.63$ & $-2.08$ & $0.72$ & $-2.95$ & $0.80$ \\
JVAS \& optical, $p_3(\lambda_0|D)$ &
$-0.77$ & $0.52$ & $-1.23$ & $0.63$ & $-1.52$ & $0.70$ & $-2.15$ & $0.79$ \\
\hline
\end{tabular*}
\end{table*}
\begin{table*}
\caption[]{Mean values and ranges for assorted confidence levels for the
parameter $\lambda_{0}$ for our a priori and various a posteriori
likelihoods from this work for $k=0$.  This should be compared to
Table~4 in Paper~I} 
\label{ta:specialk}
\begin{tabular*}{\linewidth}{@{\extracolsep{\fill}}lrrrrrrrr}
\hline
\hline
Cosmological test&
\multicolumn{2}{c}{68\% c.l.~range}  &
\multicolumn{2}{c}{90\% c.l.~range}  &
\multicolumn{2}{c}{95\% c.l.~range}  &
\multicolumn{2}{c}{99\% c.l.~range}  \\
\hline
JVAS, $p(D|\lambda_0)$ &
$-0.11$ & $0.70$ & $-0.83$ & $0.78$ & $<-1.00$ & $0.82$ & $<-1.00$ & $0.86$ \\
JVAS, $p_1(\lambda_0|D)$ &
$0.13$ & $0.75$ & $-0.15$ & $0.82$ & $-0.33$ & $0.85$ & $-0.69$ & $0.89$ \\
JVAS, $p_2(\lambda_0|D)$ &
$0.35$ & $0.77$ & $0.13$ & $0.83$ & $0.02$ & $0.85$ & $-0.21$ & $0.88$ \\
JVAS, $p_3(\lambda_0|D)$ & 
$0.41$ & $0.79$ & $0.25$ & $0.83$ & $0.16$ & $0.85$ & $-0.04$ & $0.88$ \\
JVAS \& optical, $p(D|\lambda_0)$ &
$-0.15$ & $0.45$ & $-0.49$ & $0.55$ & $-0.69$ & $0.60$ & $<-1.00$ & $0.67$ \\
JVAS \& optical, $p_1(\lambda_0|D)$ &
$0.02$ & $0.54$ & $-0.12$ & $0.61$ & $-0.29$ & $0.64$ & $-0.60$ & $0.70$ \\
JVAS \& optical, $p_2(\lambda_0|D)$ &
$0.39$ & $0.39$ & $0.09$ & $0.59$ & $0.00$ & $0.64$ & $<-0.22$ & $0.70$ \\
JVAS \& optical, $p_3(\lambda_0|D)$ &
$0.39$ & $0.51$ & $0.18$ & $0.63$ & $0.09$ & $0.66$ & $-0.09$ & $0.72$ \\
\hline
\end{tabular*}
\end{table*}

\begin{figure*}
\resizebox{0.375\textwidth}{!}{\includegraphics{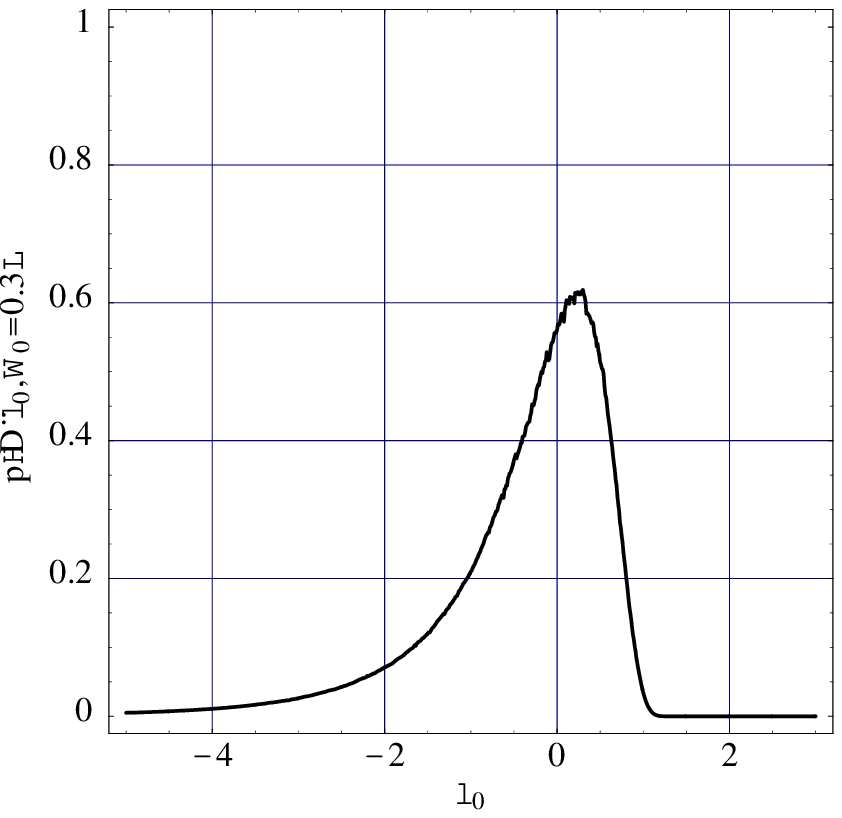}}
\hfill
\resizebox{0.375\textwidth}{!}{\includegraphics{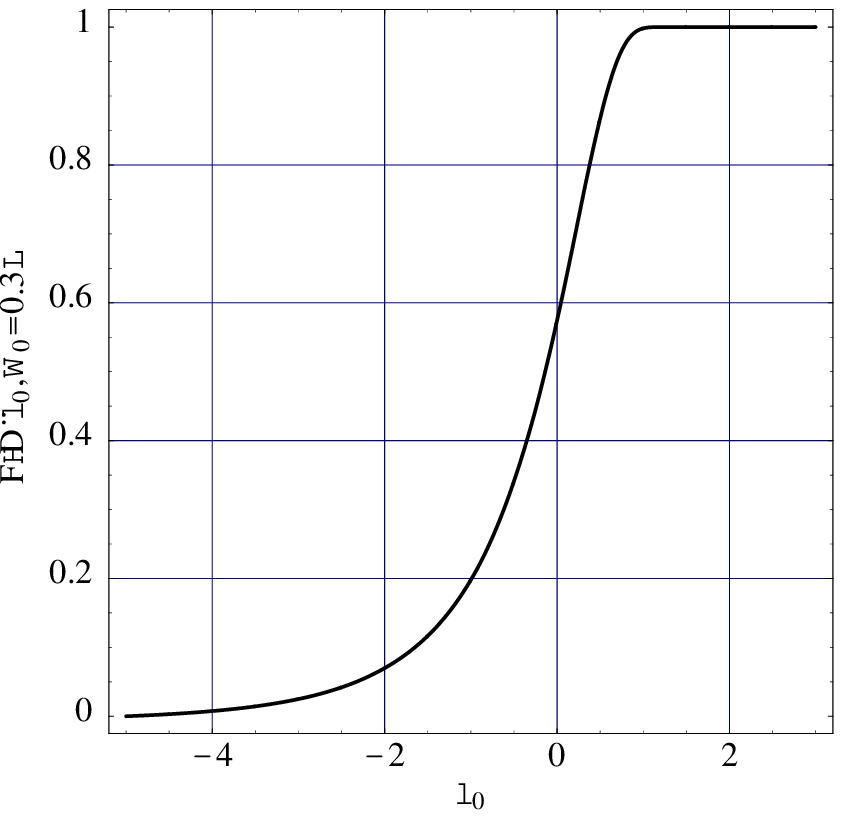}}
\caption[]{\emph{Left panel:} The likelihood function as a function of
$\lambda_{0}$ for $\Omega_{0}=0.3$ and with all nuisance parameters
taking their default values, using just the JVAS data.  \emph{Right
panel:} The same but plotted cumulatively} 
\label{fi:r03}
\end{figure*}

For a ``likely'' $\Omega_{0}$ value of 0.3 we have calculated the
likelihood with the higher resolution $\Delta\lambda_{0}=0.01$.  This is
show in Fig.~\ref{fi:r03}. 
\begin{figure*}
\resizebox{0.375\textwidth}{!}{\includegraphics{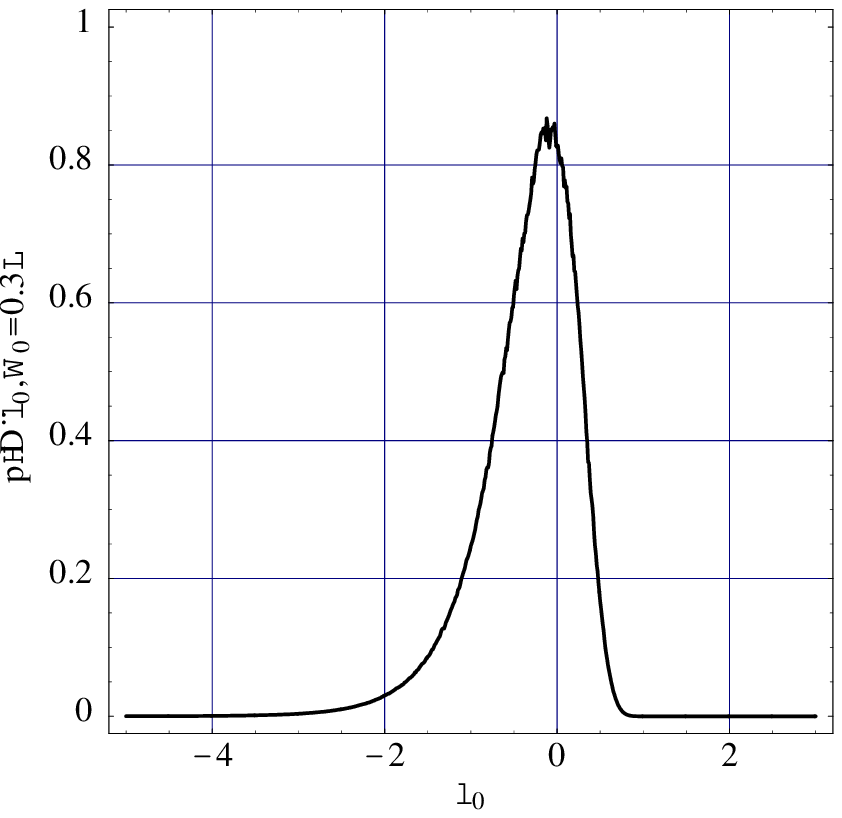}}
\hfill
\resizebox{0.375\textwidth}{!}{\includegraphics{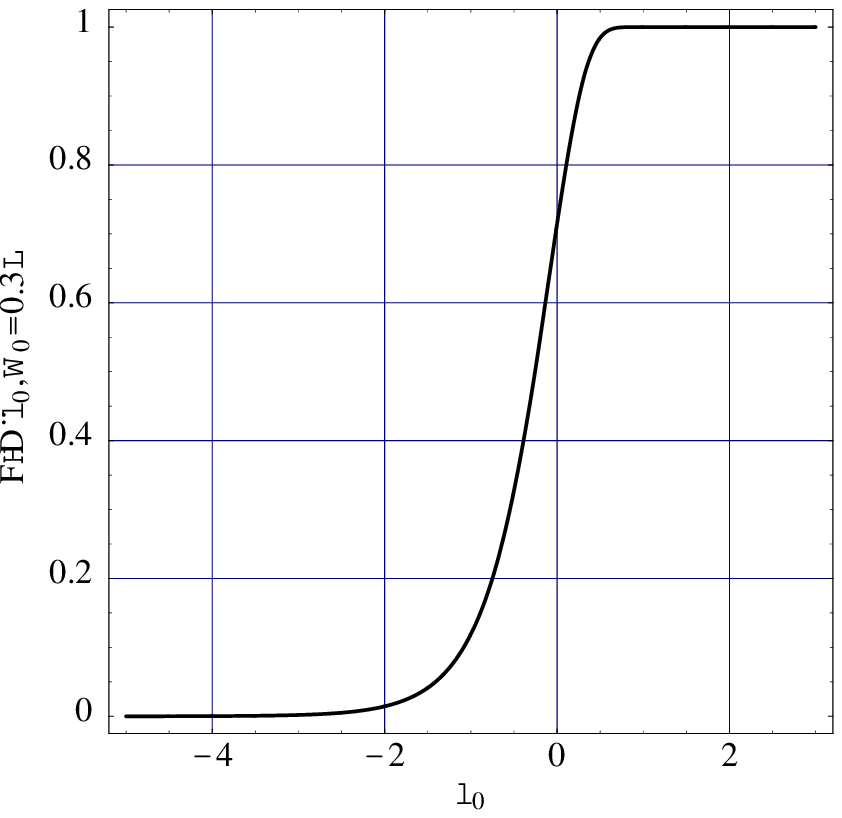}}
\caption[]{As Fig.~\ref{fi:r03} but combining optical and radio data.
\emph{Left panel:} The likelihood function as a function of
$\lambda_{0}$ for $\Omega_{0}=0.3$ and with all nuisance parameters
taking their default values.  \emph{Right panel:} The same but plotted
cumulatively} 
\label{fi:j03}
\end{figure*}
From these calculations one can extract confidence limits which, due to
the higher resolution in $\lambda_{0}$, are more accurate.  These are
presented in Table~\ref{ta:03} and should be compared to those for
$p(D|\lambda_0)$ from Table~\ref{ta:specialo}. 
\begin{table*}
\caption[]{Confidence ranges for $\lambda_{0}$ assuming
$\Omega_{0}=0.3$.  Unlike the results presented in
Table~\ref{ta:specialo}, these figures are for a specific value of
$\Omega_{0}$ and not the values of intersection of particular contours
with the $\Omega_{0}=0.3$ line in the $\lambda_{0}$-$\Omega_{0}$ plane.
These are more appropriate if one is convinced that$\Omega_{0}=0.3$ and
have been calculated using ten times better resolution than the rest of
our results presented in this work.  See Figs.~\ref{fi:r03} and
\ref{fi:j03}}
\label{ta:03}
\begin{tabular*}{\linewidth}{@{\extracolsep{\fill}}lrrrrrrrr}
\hline
\hline
data set &
\multicolumn{2}{c}{68\% c.l.~range}  &
\multicolumn{2}{c}{90\% c.l.~range}  &
\multicolumn{2}{c}{95\% c.l.~range}  &
\multicolumn{2}{c}{99\% c.l.~range}  \\
\hline
JVAS &
$-0.69$ & $0.72$ & $-1.72$ & $0.91$ & $-2.39$ & $0.98$ & $-3.83$ & $1.06$ \\
JVAS+optical &
$-0.65$ & $0.30$ & $-1.17$ & $0.49$ & $-1.48$ & $0.57$ & $-2.22$ & $0.70$ \\
\hline
\end{tabular*}
\end{table*}

As mentioned in Paper~I, to aid comparisons with other cosmological
tests, the data for the figures shown in this paper are available at 
\begin{quote}
\verb|http://multivac.jb.man.ac.uk:8000/ceres|\\
                                       \verb|/data_from_papers/JVAS.html|
\end{quote}
and we urge our colleagues to follow our example.

\section{Conclusions and outlook}
\label{conclusions}

We have used the method outlined in \citet{RQuastPHelbig99a} to
measure the cosmological constant $\lambda_{0}$ from the lensing
statistics of the Jodrell Bank-VLA Astrometric Survey.  At 95\%
confidence, our lower and upper limits on $\lambda_{0}$-$\Omega_{0}$,
using the JVAS lensing statistics information alone, are respectively
$-2.69$ and $0.68$.  For a flat universe, these correspond to lower and
upper limits on $\lambda_{0}$ of respectively $-0.85$ and $0.84$.  Using
the combination of JVAS lensing statistics and lensing statistics from
the literature as discussed in \citet{RQuastPHelbig99a} the
corresponding $\lambda_{0}-\Omega_{0}$ values are $-1.78$ and $0.27$. 
For a flat universe, these correspond to lower and upper limits on
$\lambda_{0}$ of respectively $-0.39$ and $0.64$.  Note that the lower
limit is affected more than the upper limit with respect to the
difference between the JVAS results and those in Paper~I and with
respect to combining the JVAS results with those from Paper~I. 

Our determination is consistent with other recent measurements of
$\lambda_{0}$, both from lensing statistics and from other cosmological
tests \citep[see][Paper I, for a discussion]{RQuastPHelbig99a}.  We
confirm the result of \citet[FKM]{EFalcoKM98a} that radio surveys give
higher values of $\lambda_{0}$ than optical surveys.
\citet{ACoorayQC99a} and \citet{ACooray99a} obtain a 95\% confidence upper limit on
$\lambda_{0}$ in a flat universe of 0.79 from analyses of the Hubble
Deep Field and CLASS.  However, these analyses suffer from systematic
effects due to our ignorance of the underlying flux density-dependent
redshift distribution (or, equivalently, the redshift-dependent
luminosity function) of the unlensed parent population.  As discussed in
\citet{ACooray99a}, the value of $\lambda_{0}$ obtained from CLASS will
decrease if the mean redshift of the sample is lower than presumed. 
Thus, although there is no real conflict at present as the \emph{lower}
limits on $\lambda_{0}$ are not as tight, it seems not unlikely that a
more detailed analysis of CLASS, incorporating more information about
the unlensed parent population, will result in a value more in line with
our value obtained from the JVAS analysis.  Of course, the JVAS analysis
also suffers from systematic effects, but the general agreement between
the results obtained from the analysis of optical surveys (cf.~Paper~I
and references therein) and radio surveys as presented here and in FKM
suggests that these are not overwhelming.  Also, the difference, a
higher value of $\lambda_{0}$ from radio surveys, is what one would
expect, as lens systems which go unnoticed will, all other things being
equal, reduce the value of $\lambda_{0}$.  This could be the case in
optical surveys since it is possible that extinction in the lens galaxy
and the fact that the resolution is only slightly better than the image
separation could lead to lens systems being missed.  Again, the general
agreement does suggest though that these effects are not overwhelming. 

Of course, one could imagine that the agreement is coincidental, the
optical surveys being heavily affected by extinction and resolution bias
and the radio surveys by our ignorance of the unlensed parent
population.  However, the fact that lens statistics in general gives
results which are not in conflict with other cosmological tests
(cf.~Paper~I) suggests that this is not the case.  Moreover, 
extinction would bias the results from lens
statistics and the $m$-$z$ relation (e.g.~for type Ia supernovae,
cf.~the results in Tables 3 and 4 of paper I and in the references
mentioned there) in the opposite direction.  Thus, their agreement
suggests that both methods have their systematics more or less under
control. 

The major source of uncertainty in radio lens surveys is the lack of
knowledge about the redshift distribution and number-magnitude relation
of the source sample \citep[e.g.][]{CKochanek96b}.  We are currently
undertaking the necessary observations to reduce this source of
systematic error.  Since the time scale for this project is comparable
to that for the followup of the CLASS survey, there seems little point
in doing a better analysis of JVAS in the future, especially since CLASS
is defined so that JVAS is essentially a subset of it.\footnote{The
definition of both is flat-spectrum between L-band and C-band,
i.e.~$\alpha>-0.5$ where $s_{f}\sim f^{\alpha}$, the essential
difference being the lower flux density limit of 200\,mJy for JVAS and
30\,mJy for CLASS.  However, since CLASS is defined based on newer
catalogues \citep[GB6 and NVSS:][]{PGregorySDJ96a,JCondonCGYPTB98a} than
JVAS, there will be some essentially random differences due to differing
quality of observations and variability of the sources.  All the JVAS
lenses mentioned in Table~\ref{ta:lenses} are in the new CLASS sample,
which, having no upper flux density limit, subsumes JVAS.  The previous
samples CLASS-I and CLASS-II will be similarly subsumed in the same
sense as JVAS, though the differences here will be slightly larger since
bands other than L and C were used in the preliminary definition of
these samples.} The larger size of the CLASS survey, coupled with better
knowledge of the redshift distribution and number-magnitude relation of
the source sample, should reduce both the random and systematic errors
on our value of $\lambda_{0}$.

\begin{acknowledgements}
We thank our collaborators in the JVAS, CJF and CLASS surveys for useful
discussions and for providing data in advance of publication and many
colleagues at Jodrell Bank for helpful comments and suggestions.  We
also thank John Meaburn and Anthony Holloway at the Department of
Astronomy in Manchester and the staff at Manchester Computing for
providing us with additional computational resources.  RQ is grateful to
the CERES collaboration for making possible a visit to Jodrell Bank
where this collaboration was initiated.  This research was supported in
part by the European Commission, TMR Programme, Research Network
Contract ERBFMRXCT96-0034 ``CERES''. 
\end{acknowledgements}

\bibliographystyle{aa}

\begin{thebibliography}{22}
\expandafter\ifx\csname natexlab\endcsname\relax\def\natexlab#1{#1}\fi

\bibitem[{Augusto et~al.(1999)Augusto, Browne, Wilkinson
  et~al.}]{PAugustoBWJFM99a}
Augusto P., Browne I.W.A., Wilkinson P.N., et~al., 1999, MNRAS, in preparation

\bibitem[{Biggs et~al.(1999)Biggs, Browne, Helbig et~al.}]{ABiggsBHKWP99a}
Biggs A., Browne I.W.A., Helbig P., et~al., 1999, MNRAS, 304, 349

\bibitem[{Browne et~al.(1998)Browne, Patnaik, Wilkinson \&
  Wrobel}]{IBRownePWW98a}
Browne I.W.A., Patnaik A.R., Wilkinson P.N., Wrobel J., 1998, MNRAS, 293, 257

\bibitem[{Condon et~al.(1998)Condon, Cotton, Greisen et~al.}]{JCondonCGYPTB98a}
Condon J.J., Cotton W.D., Greisen E.W., et~al., 1998, AJ, 115, 1693

\bibitem[{Cooray(1999)}]{ACooray99a}
Cooray A.R., 1999, A\&A, 342, 353

\bibitem[{Cooray et~al.(1999)Cooray, Quashnock \& Miller}]{ACoorayQC99a}
Cooray A.R., Quashnock J.M., Miller M.C., 1999, ApJ, 511, 562

\bibitem[{Falco et~al.(1998)Falco, Kochanek \& Mu{\~n}oz}]{EFalcoKM98a}
Falco E., Kochanek C.S., Mu{\~n}oz J.A., 1998, ApJ, 494, 47

\bibitem[{Fukugita et~al.(1990)Fukugita, Futamase \& Kasai}]{MFukugitaFK90a}
Fukugita M., Futamase T., Kasai M., 1990, MNRAS, 246, 24

\bibitem[{Fukugita et~al.(1992)Fukugita, Futamase, Kasai \&
  Turner}]{MFukugitaFKT92a}
Fukugita M., Futamase K., Kasai M., Turner E.L., 1992, ApJ, 393, 3

\bibitem[{Gregory et~al.(1996)Gregory, Scott, Douglas \&
  Condon}]{PGregorySDJ96a}
Gregory P.C., Scott W.K., Douglas K., Condon J.J., 1996, ApJS, 103, 427

\bibitem[{King \& Browne(1996)}]{LKingIBrowne96a}
King L.J., Browne I.W.A., 1996, MNRAS, 282, 67

\bibitem[{King et~al.(1998)King, Jackson, Blandford
  et~al.}]{LKingJBBBdBFKMNW98a}
King L.J., Jackson N.J., Blandford R.D., et~al., 1998, MNRAS, 295, L41

\bibitem[{King et~al.(1999)King, Browne, Marlow, Patnaik \&
  Wilkinson}]{LKingBMPW99a}
King L.J., Browne I.W.A., Marlow D.R., Patnaik A.R., Wilkinson P.N., 1999,
  MNRAS, in press

\bibitem[{Kochanek(1996{\natexlab{a}})}]{CKochanek96a}
Kochanek C.S., 1996{\natexlab{a}}, ApJ, 466, 638

\bibitem[{Kochanek(1996{\natexlab{b}})}]{CKochanek96b}
Kochanek C.S., 1996{\natexlab{b}}, ApJ, 473, 595

\bibitem[{Myers et~al.(1999)Myers, Rusin, Marlow et~al.}]{SMyersetal99a}
Myers S.T., Rusin D., Marlow D., et~al., 1999, in preparation

\bibitem[{Patnaik et~al.(1992)Patnaik, Browne, Wilkinson \&
  Wrobel}]{APatnaikBWW92a}
Patnaik A.R., Browne I.W.A., Wilkinson P.N., Wrobel J.M., 1992, MNRAS, 254, 655

\bibitem[{Quast \& Helbig(1999)}]{RQuastPHelbig99a}
Quast R., Helbig P., 1999, A\&A, 344, 721, in press

\bibitem[{Taylor et~al.(1996)Taylor, Vermeulen, Readhead
  et~al.}]{GTaylorVRPHW96a}
Taylor G.B., Vermeulen R.C., Readhead A.C.S., et~al., 1996, ApJS, 107, 37

\bibitem[{Turner(1990)}]{ETurner90a}
Turner E.L., 1990, ApJ, 365, L43

\bibitem[{Turner et~al.(1984)Turner, Ostriker \& Gott}]{ETurnerOG84a}
Turner E.L., Ostriker J.P., Gott III J.R., 1984, ApJ, 284, 1

\bibitem[{Wilkinson et~al.(1998)Wilkinson, Browne, Patnaik, Wrobel \&
  Sorothia}]{PWilkinsonBPWS98a}
Wilkinson P.N., Browne I.W.A., Patnaik A.R., Wrobel J., Sorothia B., 1998,
  MNRAS, 300, 790

\end{thebibliography}

\end{document}